\documentclass[aps,superscriptaddress,twocolumn,twoside,floatfix,prx,nofootinbib,a4paper]{revtex4-2}
\pdfoutput=1

\usepackage[UKenglish]{babel}
\usepackage[UKenglish,cleanlook]{isodate}   
\usepackage{times}
\usepackage{epsfig}
\usepackage{amsfonts}
\usepackage{amsmath}
\usepackage{amssymb}
\usepackage{IEEEtrantools}
\usepackage{color}
\usepackage{multirow}
\usepackage[normalem]{ulem}
\usepackage{latexsym}
\usepackage{IEEEtrantools}
\usepackage{amsfonts}
\usepackage{mathrsfs}
\usepackage{natbib}
\usepackage{verbatim}
\usepackage[T1]{fontenc}
\usepackage{amsthm}
\usepackage{tabularx,ragged2e}
\usepackage{array,booktabs}
\newcolumntype{C}{>{$}c<{$}}
\AtBeginDocument{
	\heavyrulewidth=.08em
	\lightrulewidth=.05em
	\cmidrulewidth=.03em
	\belowrulesep=.65ex
	\belowbottomsep=0pt
	\aboverulesep=.4ex
	\abovetopsep=0pt
	\cmidrulesep=\doublerulesep
	\cmidrulekern=.5em
	\defaultaddspace=.5em
}

\usepackage{graphicx}
\usepackage{subfig}
\usepackage{caption}
\usepackage[colorlinks=true,linkcolor=blue,citecolor=magenta,urlcolor=blue]{hyperref}

\DeclareMathOperator{\Tr}{tr}

\definecolor{darkgreen}{rgb}{0,0.5,0}

\newcommand{\ket}[1]{|#1\rangle}
\newcommand{\bra}[1]{\langle#1|}
\newcommand{\braket}[2]{\langle#1|#2\rangle}

\newcommand{\ketbra}[2]{\ket{#1}\bra{#2}}

\newcommand{\norm}[2][]{#1\lVert #2 #1\rVert}
\newcommand{\abs}[1]{\lvert #1 \rvert}

\newcommand{\Babs}[1]{\Bigl\lvert #1 \Bigr\rvert}

\newtheorem*{lemma1}{Lemma 1}
\newtheorem{definition}{Definition}
\newtheorem{proposition}{Proposition}

\newtheorem*{def1b}{Definition 1'}
\newtheorem*{def2b}{Definition 2'}

\newcommand{\ba}{\begin{eqnarray}}
\newcommand{\be}{\begin{equation}}
\newcommand{\ee}{\end{equation}}

\newcommand{\ea}{\end{eqnarray}}
\newcommand{\ban}{\begin{eqnarray*}}
	\newcommand{\ean}{\end{eqnarray*}}

\definecolor{orange}{rgb}{1,0.5,0}

\begin{document}
	

\title{Correlations in entanglement-assisted prepare-and-measure scenarios}
	

	\author{Armin Tavakoli}\thanks{These authors share the first authorship.}
	\affiliation{Institute for Quantum Optics and Quantum Information - IQOQI Vienna, Austrian Academy of Sciences, Boltzmanngasse 3, 1090 Vienna, Austria}
	\affiliation{D\'epartement de Physique Appliqu\'ee, Universit\'e de Gen\`eve, CH-1211 Gen\`eve, Switzerland}	
	
	\author{Jef Pauwels}\thanks{These authors share the first authorship.}
	\affiliation{Laboratoire d'Information Quantique, Universit\'e libre de Bruxelles (ULB), Belgium}

	\author{Erik Woodhead}
	\affiliation{Laboratoire d'Information Quantique, Universit\'e libre de Bruxelles (ULB), Belgium}
	
	\author{Stefano Pironio}
	\affiliation{Laboratoire d'Information Quantique, Universit\'e libre de Bruxelles (ULB), Belgium}
	
	\date{September~26th,~2021}

\begin{abstract}
		
We investigate the correlations that can arise between Alice and Bob in prepare-and-measure communication scenarios where the source (Alice) and the measurement device (Bob) can share prior entanglement. The paradigmatic example of such a scenario is the quantum dense coding protocol, where the communication capacity of a qudit can be doubled if a two-qudit entangled state is shared between Alice and Bob. We provide examples of correlations that actually require more general protocols based on higher-dimensional entangled states.  
This motivates us to investigate the set of correlations that can be obtained from communicating either a classical or a quantum $d$-dimensional system in the presence of an unlimited amount of entanglement. We show how such correlations can be characterized by a hierarchy of semidefinite programming relaxations by reducing the problem to a non-commutative polynomial optimization problem. We also introduce an alternative relaxation hierarchy based on the notion of informationally-restricted quantum correlations, which, though it represents a strict (non-converging) relaxation scheme, is less computationally demanding. 
 As an application, we introduce device-independent tests of the dimension of classical and quantum systems that, in contrast to previous results, do not make the implicit assumption that Alice and Bob share no entanglement. We also establish several relations between communication with and without entanglement as resources for creating correlations.
\end{abstract}
	
	\maketitle
	

\section{Introduction}
\begin{figure}[t!]
	\centering
	\subfloat[Shared randomness.]{
\includegraphics{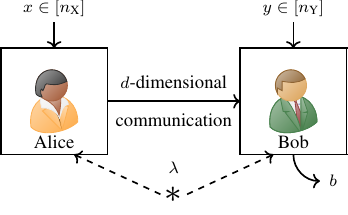}}
	\\
	\subfloat[Shared entanglement.]{
\includegraphics{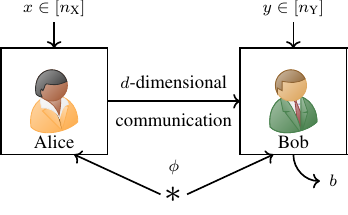}}
	\caption{Prepare-and-measure scenario. Alice encodes an input $x$ into a physical system communicated to Bob. Bob measures the incoming system depending on an input $y$ and obtains an output $b$. We are interested in characterizing the possible conditional probabilities $p(b|x,y)$ if the communication is limited to $d$-dimensional messages. Much of past research has considered the case (a) where Alice and Bob are initially independent or  share classical randomness. We consider the situation (b) where they share quantum entanglement.\label{FigScenario} }
\end{figure}

The archetype communication scenario, ubiquitous in classical and quantum information theory, is the prepare-and-measure scenario illustrated in Figure \ref{FigScenario}a. Alice prepares a physical system, depending on some input $x\in\{1,\ldots, n_\text{X}\}$, and sends it to Bob. Bob then performs on the incoming system a measurement, according to some choice of input $y\in\{1,\ldots,n_\text{Y}\}$, and obtains an output $b\in\{1,\ldots,n_\text{B}\}$. From an operational perspective, this prepare-and-measure scenario is completely characterized by the conditional probabilities $p(b|x,y)$, which describe the correlations that are established between Alice and Bob. These correlations are limited by the amount of communication carried by the physical systems from Alice to Bob.

 Communication may naturally be, and is commonly, quantified in terms of the dimension $d$ of the exchanged messages, i.e, the alphabet size for classical messages and the dimension of the Hilbert space for quantum messages. Consequently, much research has been directed at studying the correlations $p(b|x,y)$ that arise from the communication of a classical or quantum $d$-dimensional system. This covers a wide range of topics including foundations of quantum theory \cite{InfoCausality, Brassard2006}, dimension witnessing \cite{Wehner2008, Gallego2010, Brunner2013, Vicente2017}, random access coding \cite{Ambainis1999, Ambainis2006,  Tavakoli2015}, quantum random number generation \cite{Li2011, Li2012}, quantum key distribution \cite{Pawlowski2011B, Woodhead2015}, self-testing \cite{Tavakoli2018, Farkas2019, Tavakoli2020} and various protocols for characterising and certifying quantum devices \cite{FakeTriangle, Mironowicz2019, Tavakoli2020B}. It has also motivated a considerable number of experiments  (see e.g.~\cite{Trojek2005, Ahrens2012, Hendrych2012, Muhammad2014, Ambrosio2014, Smania2016, Martinez2018}). 

Typically, quantum communication models, e.g.~as in the references above,  consider Alice and Bob as initially independent or allow them to share  a classical random variable (as in Figure~\ref{FigScenario}a). However, quantum theory naturally enables a more general communication scenario in which Alice and Bob share prior entanglement (as in Figure~\ref{FigScenario}b). The introduction of entanglement to assist classical and quantum communication should enlarge the set of possible correlations between Alice and Bob. Indeed, while entanglement itself cannot be used for communication, it is well-known to amplify the capacity of quantum channels \cite{Bennett2002}, most famously via the quantum dense coding protocol \cite{DenseCoding}. In fact, even if Alice only communicates classical messages to Bob, prior entanglement provides an advantage for different tasks, such as communication complexity \cite{Buhrman2001, Brukner2003,Tavakoli2017} and random access codes \cite{Pawlowski2010}.

Understanding how the presence of entanglement impacts the set of possible correlations between Alice and Bob is also important for analyzing the security of semi-device-independent prepare-and-measure protocols such as random number generation and quantum key distribution. Commonly, such protocols are based on unentangled devices. Nevertheless, even if Alice's and Bob's devices are initially uncorrelated, quantum messages in the early communication rounds  could be used to build up shared entanglement that could then be exploited in later rounds in order to corrupt the protocol. 

In spite of the conceptual and practical interest, much less is known about prepare-and-measure scenarios with entanglement than scenarios without entanglement.  Particularly noteworthy is that, in contrast to the setting without entanglement \cite{Navascues2015, TavakoliSymmetry}, no general technique is known for bounding (from the exterior) the set of correlations $p(b|x,y)$ that can be generated by $d$-dimensional messages assisted by a, potentially unbounded amount of, entanglement. In this work, we address this central question and initiate a systematic study of prepare-and-measure scenarios with entanglement.

In section~\ref{sec:def} we  define formally the entanglement-assisted (EA) communication scenario that we consider. The quantum dense coding protocol, which is the paradigmatic example by which entanglement can enhance quantum communication, exploits an entangled pair of the same local dimension as the quantum communication (i.e., an entangled qubit pair in the case that a qubit is transmitted). In section~\ref{sec:beyonddc} we show that certain correlations that can be achieved by sending an EA qubit require higher-dimensional entanglement. The classical analog of this result is established in \cite{companionpaper}. Motivated by these observations, we proceed in section~\ref{sec:methods} by addressing the general question of characterizing the set of correlations achievable with $d$-dimensional classical and quantum communication when the communicating parties may share any amount of entanglement. We connect this problem to non-commutative polynomial optimization \cite{NCpoly} and to the recently developed concept of informationally restricted correlations \cite{InfoCorrelations, InfoCorrelations2}. This allows us to bound the correlations using a hierarchy of semidefinite programming (SDP) relaxations. In section~\ref{sec:applications}, we apply  our methods to different device-independent tests of classical and quantum dimension. In all considered examples, our method produces either verifiably optimal bounds or (at worst) nearly-optimal bounds. Our more general setting leads us to re-examine the conclusions one can draw from such dimension tests in light of shared entanglement. Finally, in section~\ref{sec:resources} we apply our methods to investigate the relationship between entanglement and quantum communication as resources for creating correlations. We show that there exists situations where either resource can outperform the other. While such questions have been the topic also of  previous research efforts \cite{Pawlowski2010, Pawlowski2012, Tavakoli2016, Hameedi2017, Tavakoli2017b, Martinez2018}, our analysis requires no additional assumptions and is tolerant to noise.

\section{Correlations from entanglement-assisted $d$-dimensional communication}\label{sec:def}
Consider an experiment featuring two parties, Alice and Bob, who share an arbitrary, and without loss of generality, pure entangled state $\ket{\phi_{AB}}\in \mathcal{H}_A\otimes \mathcal{H}_B$\footnote{The requirement that the shared state be pure is not restrictive because we do not limit the dimension of the Hilbert space $\mathcal{H}_B$. This means that, even if the physical state is mixed, it can always be purified i.e. we can understand every mixed state $\rho \in \mathcal{H}_A \otimes \mathcal{H}_{B_0}$ as the partial trace by Bob of a pure state $\ket{\phi} \in \mathcal{H}_A \otimes \mathcal{H}_{B_0B_1}$.}. Alice receives an input $x$ from the set $[n_\text{X}]\equiv \{1,\ldots,n_\text{X}\}$ and encodes her input, possibly using her share of the entangled state $\ket{\phi}$, into a system $C$ of dimension no greater than $d$ that is sent to Bob. Bob receives an input $y\in[n_\text{Y}]\equiv \{1,\ldots,n_\text{Y}\}$ and performs a measurement, depending on $y$, on the incoming system $C$ and his share of the entangled state. The outcome of this measurement is denoted $b\in[n_\text{B}]\equiv \{1,\ldots,n_\text{B}\}$. This scenario is characterized by the conditional probability distributions $p(b|x,y)$, which we refer to as the correlations.

\begin{figure}[t!]
	\centering
	\subfloat[General channel.]{
\includegraphics{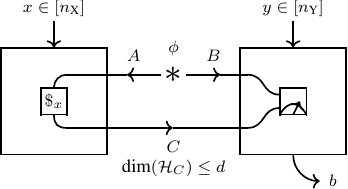}}
\\
\subfloat[qc-channel.]{
\includegraphics{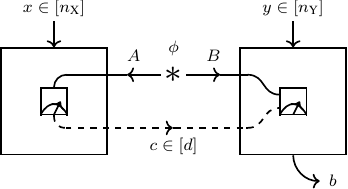}}
\caption{(a) When Alice and Bob share an entangled state $\phi$, Alice encodes her classical input $x$ into a $d$-dimensional message $c$ by applying a  quantum channel $\$_x$. Depending on his classical input $y$, Bob performs a joint measurement $\{M_{b|y}\}_b$ on Alice's message and his share of the entanglement and produces an outcome $b$. (b) When the communication is classical, Alice performs a POVM $\{M_{c \vert x}\}_c$ and relays the outcome $c$ to Bob. Bob's operation can be viewed as a POVM $\{M_{b\vert y,c} \}_{y,c}$ conditioned on his classical input $y$ and Alice's message $c$. \label{FigPMsharedEnt}}
\end{figure}

The most general way that Alice can exploit her share of the entangled state when encoding her classical input $x$ into the $d$-dimensional system $C$ is through the application of a completely positive trace-preserving (CPTP) map $\$_x \colon L(\mathcal{H}_A) \to L(\mathcal{H}_C)$ from the space $L(\mathcal{H}_A)$ of linear operators on $\mathcal{H}_A$ to the space $L(\mathcal{H}_C)$ of linear operators on $\mathcal{H}_C\simeq\mathbb{C}^d$. The total state available to Bob, composed of the communicated $d$-dimensional quantum system $C$ from Alice and of his share of $\ket{\phi}$, is then $\tau_{CB}^x\equiv \left(\$_x\otimes \openone_B\right) [\ketbra{\phi_{AB}}{\phi_{AB}}]$. The most general measurement he can perform on this state when selecting input $y$ is then given by a measurement (POVM) with elements $\{M_{b|y}\}_b$. This is illustrated in Figure \ref{FigPMsharedEnt}a and leads to the following definition.
\begin{definition}
	We say that the correlations $p(b|x,y)$ can be reproduced by an EA $d$-dimensional quantum communication protocol if there exists
	\begin{itemize}
		\item a bipartite pure entangled state $\ket{\phi_{AB}}$ in $\mathcal{H}_A\otimes \mathcal{H}_B$,    where $A$ and $B$ are physical systems with finite or separable\footnote{A Hilbert space is separable iff it admits a countable orthonormal basis.} Hilbert spaces $\mathcal{H}_A$ and $\mathcal{H}_B$,
		\item a CPTP map $\$_x \colon L(\mathcal{H}_A) \to L(\mathcal{H}_C)$ from system $A$ to a system $C$ with a $d$-dimensional Hilbert space $\mathcal{H}_C \simeq \mathbb{C}^d$ for each input $x$,
		\item an $n_\text{B}$-outcome POVM $\{M_{b|y}\}_b$ on the joint systems $C$ and $B$ for each input $y$,  
	\end{itemize}
	such that
	\begin{equation}
		p(b|x,y)=\Tr\left(\tau_{CB}^x\, M_{b|y}\right),
	\end{equation}	
	where 
	\begin{equation}\label{totalstate}
		\tau_{{CB}}^x\equiv \left(\$_x\otimes \openone_B\right) \bigl[\ketbra{\phi_{AB}}{\phi_{AB}}\bigr]\,.
	\end{equation}
\end{definition}

The above definition is fully general and assumes that the communication from Alice to Bob is quantum. However, we can also restrict the communication to be classical. This can be represented in the Hilbert space formalism of quantum theory by imposing that the CPTP maps $\$_x$ output diagonal, classical states: $\$_x[\rho]=\sum_{c=1}^d p(c|x,\rho)\,\ketbra{c}{c}$ for all $\rho\in L(\mathcal{H}_A)$. The Riesz representation theorem asserts that, for every $x$, linear maps of the form $p(c|x,\rho)$ can be written in terms of the Born-rule.  Therefore, any CPTP map of this form represents the outcome of a POVM $\{M_{c|x}\}_c$ performed on the input state $\rho$: $\$_x[\rho]=\sum_{c=1}^d \Tr\left(\rho\, M_{c|x}\right)\,\ketbra{c}{c}$. The states available to Bob are then the classical-quantum (cq) states $\tau_{{CB}}^x = \left(\$_x\otimes \openone_B\right) [\ketbra{\phi_{AB}}{\phi_{AB}}] = \sum_{c=1}^d \ketbra{c}{c}\otimes \tau_{B}^{c,x}$ where $\tau_{B}^{c,x}=\Tr_A\left(\ketbra{\phi_{AB}}{\phi_{AB}}\, M_{c|x}\otimes \openone_B\right)$ is the (subnormalized) reduced state of Bob when Alice performs the POVM $\{M_{c|x}\}$ on her share of $\ket{\phi_{AB}}$ and gets outcome $c$. Any measurement by Bob on such a cq-state can be viewed as Bob first reading the classical system $C$ and then performing a measurement on the quantum system $B$ depending on the value $c$ he obtained. This is illustrated in Figure \ref{FigPMsharedEnt}b.  The correlations Alice and Bob generate are then $p(b|x,y)=\sum_c \Tr\left(\tau_{B}^{c,x}\,M_{b|y,c} \right)=\sum_c \Tr\left(\ketbra{\phi_{AB}}{\phi_{AB}}\,M_{c|x}\otimes M_{b|y,c}\right)$. We thus have the following definition in the classical case.
\begin{definition}
	We say that the correlations $p(b|x,y)$ can be reproduced by an EA $d$-dimensional classical communication protocol if there exists 
	\begin{itemize}
		\item a bipartite pure entangled state $\ket{\phi_{AB}}$ in $\mathcal{H}_A\otimes \mathcal{H}_B$, where $A$ and $B$ are physical systems with finite or separable Hilbert spaces $\mathcal{H}_A$ and $\mathcal{H}_B$
		\item a $d$-outcome POVM $\{M_{c|x}\}_c$ on $A$ for each input $x$, 
		\item an $n_{\text{B}}$-outcome POVM $\{M_{b|y,c}\}_b$ on $B$ for each input $y$ and $c\in [d]$,  
	\end{itemize}
	such that
	\begin{equation}\label{eq:corrclass}
		p(b|x,y)=\sum_{c=1}^d \Tr\left(\ketbra{\phi_{AB}}{\phi_{AB}}\, M_{c|x}\otimes M_{b|y,c}\right).
	\end{equation}
\end{definition}
Note that \eqref{eq:corrclass} simply represents Bob's marginal correlations in a kind of bipartite Bell experiment were the measurement performed on Bob's side depends not only on his input $y$ but also on the communicated output $c$ of Alice's measurement.

\section{Beyond dense coding: qubit communication enhanced by four-dimensional entanglement}\label{sec:beyonddc}

The simplest form of quantum communication has Alice sending qubits ($d=2$) to Bob. While a single qubit can only carry only one bit of information \cite{Holevo}, it is well known that if Alice and Bob share the maximally entangled two-qubit state 
\begin{equation}\label{phimax}
  \ket{\phi_\text{max}}=\frac{1}{\sqrt{2}}\bigl(\ket{00}+\ket{11}\bigr),
\end{equation}
Alice can use a single qubit of communication to send 2 bits of information to Bob; this is the quantum dense coding protocol \cite{DenseCoding}. Specifically, in this celebrated protocol,  Alice has four possible inputs $x=(x_1,x_2)\in\{0,1\}^2$ and, given $x$, applies the Pauli unitary $X^{x_2}Z^{x_1}$ to her share of $\ket{\phi_\text{max}}$ before sending it to Bob. Bob's total state $\ket{\tau_{CB}^x}=\left(X^{x_2}Z^{x_1}\otimes \openone_B\right)\ket{\phi_\text{max}} $ then corresponds to one the four Bell-states $\{(\ket{00}\pm\ket{11})/\sqrt{2},(\ket{01}\pm\ket{10})/\sqrt{2}\}$ depending on Alice's input $x$. Since these states form a basis, Bob can deterministically learn the value of $x$ by measuring in this basis, thus allowing Alice to send two bits to Bob. More generally, this protocol enables Bob to generate any correlations $p(b|x,y)$ in a protocol with $n_\text{X}=4$, since knowing Alice's input $x$ and his input $y$, Bob can sample $b$ according to the desired distribution $p(b|x,y)$.

Note that from the perspective of the general definition introduced in the previous section, the dense coding protocol is particular in that the shared entangled state is of the same local dimension as the communicated quantum system and the CPTP maps applied by Alice are unitaries. We now provide a qubit communication example, based on a modified random access coding task, where entanglement of local dimension \emph{four} processed by non-unitary CPTP maps outperforms any strategy based on two-dimensional entanglement.

\subsection{Random Access Code with flagged input}\label{sec:flag}
The starting point for the task that we introduce is the usual $2 \rightarrow 1$ quantum random access code (RAC) \cite{Ambainis1999}, where Alice must encode two bits $x=(x_1,x_2)\in\{00,01,10,11\}$ in a single qubit such that Bob is able to guess as best as possible either the first bit, if $y=1$, or the second bit, if $y=2$. Denoting Bob's guess $b\in\{0,1\}$ and assuming that Alice and Bob's inputs are chosen uniformly, the success probability of Bob is given by $\frac{1}{8}\sum_{x_1,x_2=0}^1\sum_{y=1}^2 p(b=x_y|(x_1,x_2),y)$.

For convenience, we introduce the change of notation $x\in\{00,01,10,11\}\rightarrow \{1,2,3,4\}$ and $b\in\{0,1\}\rightarrow b\in\{1,-1\}$. We can then write the success probability as $1/2+W_\text{RAC}/16$ where $W_{\text{RAC}}$ is the RAC correlation function
\begin{equation}\label{Wexp}
    W_\text{RAC}=\sum_{x=1}^4\sum_{y=1}^2 c_{xy}E_{xy},
    \end{equation}
    $E_{xy}=p(1|x,y)-p(-1|x,y)$ denotes the expectation value of $b$ and the $4\times 2$ coefficients $c_{xy}$  are given by 
    \begin{align}\label{coefs}
    & c=\begin{pmatrix}
    1 & 1 \\
    1 & -1  \\
    -1 & 1 \\
    -1 & -1 
    \end{pmatrix} \,.
    \end{align}

Obviously, if shared entanglement is present then a value of $W_\text{RAC}=8$ (corresponding to a success probability of 1) is possible, as Alice can perfectly encode her 4 possible inputs into a single qubit using the dense coding protocol. To make the task non-trivial we add the following modification. We assume that Alice has a fifth possible choice of input $x=5$ and Bob has a third input $y=3$. This additional input of Alice can be thought of as a special, flagged, input (e.g. that communicates a very important or urgent matter) which must be unambiguously identified by Bob whenever he decides on $y=3$. This can be represented by adding the following constraints to our task
\begin{align}\label{constraint}
    & E_{13}=E_{23}=E_{33}=E_{43}=-E_{53}=1\,,
\end{align}
i.e., when Bob uses input $y=3$, he must necessarily get output $b=1$ if $x=1,2,3,4$ and $b=-1$ if $x=5$, allowing him to identify perfectly whether $x=5$ was sent or not. 

In summary, our scenario corresponds to $n_\text{X}=5$, $n_\text{Y}=3$, $n_\text{B}=2$  and we are interested in the maximal value of \eqref{Wexp} subject to the constraints \eqref{constraint} when Alice communicates a quantum system to Bob of dimension $d=2$.

Clearly we can not achieve $W_\text{RAC}=8$ while respecting the constraint \eqref{constraint} as this would imply that Bob can perfectly guess the five inputs of Alice, i.e., that Alice can communicate to Bob $\log_2(5)$ bits, while we recall that an EA qubit only can carry at most two bits of information. 

A strategy directly based on the dense coding protocol can achieve a value $W_\text{RAC}=6$. Indeed, it amounts to a strategy where two classical bits are sent from Alice to Bob. But since the input $x=5$ must be perfectly discriminated from the inputs $x=1,\ldots,4$, this means that effectively Alice encodes the four inputs $x=1,\ldots,4$ using a classical trit. The best value of the $2\rightarrow 1$ RAC function \eqref{Wexp} when communicating a trit is known to be $6$ \cite{Ahrens2014}. We show in the next subsection that there are strategies using a two-qubit entangled state that are more effective than the dense coding protocol and in the next one that strategies based on two-ququart entanglement are even better.

\subsection{Strategies based on two-dimensional entanglement}\label{sec:twodim}
Consider the following simple strategy for evaluating $W_\text{RAC}$ under the constraint \eqref{constraint} when the entanglement is restricted to a two-qubit state. The intuition stems directly from the quantum dense coding protocol. Let Alice and Bob share the maximally entangled state \eqref{phimax} and let Alice, on her share of the state, apply the unitaries $((\openone-i\sigma_x)/\sqrt{2},\openone,\sigma_x,\sigma_y,\sigma_z)$ for inputs, respectively, $x=1,2,3,4,5$. She then communicates the transformed qubit to Bob. It is immediate that the states $\tau_{CB}^x$ for $x=1,\ldots,4$ live in the subspace $\{(\ket{00}+\ket{11})/\sqrt{2},(\ket{01}\pm\ket{10})/\sqrt{2}\}$ while $\ket{\tau_{CB}^5} = \ket{00}-\ket{11})/\sqrt{2}$ is in the orthogonal complement. Thus the input $x=5$ can be completely discriminated from the other inputs using an appropriate measurement for $y=3$ and the constraint \eqref{constraint} is satisfied. Replacing $E_{xy} = \Tr\left(\tau_{CB}^x\,M_y\right)$ in \eqref{Wexp} where $M_y=M_{1|y}-M_{-1|y}$ is the observable associated to Bob's input $y$, we have
\begin{IEEEeqnarray}{rCl}
  W_\text{RAC} &=& \Tr \bigl( (\tau_{CB}^1+\tau_{CB}^2-\tau_{CB}^3-\tau_{CB}^4) M_1 \bigr)
  \nonumber \\
  &&+\> \Tr \bigl( (\tau_{CB}^1-\tau_{CB}^2+\tau_{CB}^3-\tau_{CB}^4) M_2 \bigr) \,,
\end{IEEEeqnarray}
which is maximized when the $\pm 1$ eigenspace of $M_y$ is the $\pm 1$ eigenspace of the combination of states appearing in the traces. This leads to 
\begin{IEEEeqnarray}{rCl}
  \label{res}
  W_\text{RAC} &=& \Tr \bigl(\abs{\tau_{CB}^1+\tau_{CB}^2-\tau_{CB}^3-\tau_{CB}^4}\bigr)
  \nonumber \\
  &&+\> \Tr \bigl(\abs{\tau_{CB}^1-\tau_{CB}^2+\tau_{CB}^3-\tau_{CB}^4}\bigr)
  \nonumber \\
  &=& 2 (1+\sqrt{5}) \nonumber \\
  &\approx& 6.47
\end{IEEEeqnarray}
for the specific states chosen above. This strategy thus makes a better use of the shared entanglement than one directly based on the dense coding protocol.

It turns out that no larger value of $W_\text{RAC}$ is possible by means of qubit communication assisted by two-qubit entanglement. To prove this, note that the states $\tau_{CB}^x$ in Bob's possession are four dimensional since $\mathcal{H}_C\simeq \mathbb{C}^2$ (Alice communicates a qubit) and $\mathcal{H}_B\simeq \mathbb{C}^2$ (we assume the shared entanglement is of local dimension two). These states actually occupy a subset of the total four dimensional Hilbert space since they must satisfy condition \eqref{constraint}. Let us relax this condition and consider the more generous situation where the states $\tau_{CB}^x$ live in an unconstrained four-dimensional space. This does not decrease the largest possible value of $W_\text{RAC}$ and it simplifies the analysis of the problem. The constraint \eqref{constraint} implies that Alice's first four states must be confined to a three-dimensional Hilbert space orthogonal to her fifth state. This reduces the problem to one of evaluating the largest value of the RAC function $W_\text{RAC}$ when the four relevant states ($x\in\{1,\ldots,4\}$) are encoded in a qutrit. This problem has been addressed in previous literature \cite{NavascuesLong, Tavakoli2018} where it was shown that the optimal quantum implementation achieves the value given in \eqref{res}. Actually, the strategy that we described above is a straightforward reformulation of this optimal qutrit strategy to our EA qubit scenario.

\subsection{Strategy based on four-dimensional entanglement}\label{sec:fourdim}
We now show that qubit communication assisted by higher-dimensional entanglement can further improve the value of $W_\text{RAC}$. Specifically, we show that a value of $W_\text{RAC}$ larger than that in \eqref{res} is possible if Alice and Bob share two copies of the maximally entangled two-qubit state: $\ket{\phi}_{AB}=\ket{\phi_\text{max}}_{A_1B_1}\otimes \ket{\phi_\text{max}}_{A_2B_2}$. Consequently, the states held by Bob after Alice's communication are of dimension eight, corresponding to a qubit system (the communication) and the ququart system (Bob's share of $\ket{\phi}_{AB}$). 

Alice's strategy consists in applying a two-qubit unitary $U_x$ on her systems $A_1A_2$, binning the first qubit $A_1$ and sending the second qubit $A_2$ to Bob. The channel she implements is thus given by 
\begin{equation}
\includegraphics{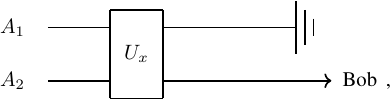}
 \label{eq:q2E}
\end{equation}

where the unitary operations are given by
\begin{align}
    U_1 &= \openone \otimes \openone,
    \label{eq:u1} \\
	U_2 &= CNOT_1 \, CNOT_2, \\
	U_3 &= \openone \otimes \sigma_x \, CNOT_1 \, CNOT_2, \\
	U_4 &= \openone \otimes \sigma_z, \label{eq:u4} \\
	U_5 &= \openone \otimes \sigma_z\sigma_x \, CNOT_2,
\end{align}
and $CNOT_i$ is the controlled-$NOT$ gate with the control on the $i$th qubit.

It can be checked that the total states $\tau_{CB}^x$ Bob measures in his laboratory are then rank-2 states of the form
\begin{equation}
\tau_{CB}^x=\frac{1}{2}\bigl(\ketbra{\psi_x}{\psi_x}+\ketbra{\varphi_x}{\varphi_x}\bigr)
\end{equation}
where the states $\ket{\psi_x}$ and $\ket{\phi_x}$ are readily computed from the unitaries given above. In particular one finds that $\ket{\psi_5}$ and $\ket{\phi_5}$ are orthogonal to all other states and thus that the constraint \eqref{constraint} is satisfied through a proper choice of measurement for $y=3$. The optimization of Bob's measurements for $y=1,2$ can be performed as in the previous subsection and, replacing the specific states obtained from the unitaries \eqref{eq:u1}-\eqref{eq:u4} in Eq.~\eqref{res}, we obtain
\begin{equation}\label{res2}
W = 2(2+\sqrt{2}) \approx 6.83 \,,
\end{equation}
which exceeds the bound $W_\text{RAC}\leq 6.47$ for qubit communication assisted by two-qubit entanglement. 

In summary, we have shown in this section that in EA $d$-dimensional quantum communication protocols we cannot restrict the entanglement to be of local dimension $d$. We establish a similar result for the case of EA classical communication in \cite{companionpaper}. Whether some finite upper-bound on the entanglement dimension can generally be assumed is an interesting question not resolved here. The analogy to the usual Bell scenario would suggest a negative answer \cite{Coladangelo}.

\section{Semidefinite programs for correlations in entanglement-assisted $d$-dimensional communication}\label{sec:methods}

We now describe how to characterize through sequences of semidefinite programming (SDP) approximations the set of correlations achievable from $d$-dimensional quantum or classical communication. We consider both inner and outer characterizations that approximate the quantum set from the inside and the outside.

\subsection{Inner characterization through seesaw iterations}

In definition~1, the correlations $p(b|x,y)$ are expressed as the result of a measurement performed by Bob on a state $\tau_{{CB}}^x$ resulting from Alice's application of a CPTP map on her part of an entangled state shared with Bob. 
This representation can be simplified using state-channel duality \cite{Wilde}\footnote{This simplification was also recently noticed in \cite{moreno}.}. Specifically, by exploiting the isomorphism between CPTP maps and quantum states, we may represent the total state $\tau_{{CB}}^x$ as a (generally mixed) bipartite state in $L\left(\mathcal{H}_{C}\otimes \mathcal{H}_B\right)$ with the property that the marginal state on system $B$ is independent of Alice's input $x$ (no-signaling): $\Tr_\text{C}\left(\tau_{{CB}}^x\right)=\tau_{B}$ for all $x$. 
We thus have the following definition equivalent to definition~1
\begin{def1b}
	We say that the correlations $p(b|x,y)$ can be reproduced by an EA $d$-dimensional quantum communication protocol if there exists
	\begin{itemize}
        \item a bipartite entangled state $\tau_{{CB}}^x \in L\left(\mathcal{H}_{C}\otimes \mathcal{H}_B\right)$, where $C$ is a physical system with $d$-dimensional Hilbert space $\mathcal{H}_C\simeq \mathbb{C}^d$ and $B$ is a physical system with finite or separable Hilbert space $\mathcal{H}_B$, for each input $x$, where the states $\tau_{{CB}}^x$ all have the same marginal state $\tau_B$:
		\begin{equation}\label{marginal}
			\Tr_\text{C}\left(\tau_{{CB}}^x\right)=\tau_{B}\quad\text{  for all } x,
		\end{equation}
		\item an $n_{B}$-outcome POVM $\{M_{b|y}\}_b$ on the joint sytems $C$ and $B$ for each input $y$, 
	\end{itemize}
	such that
	\begin{equation}
		p(b|x,y)=\Tr\left(\tau_{{CB}}^x\,M_{b|y}\right).
	\end{equation}
\end{def1b}
In the case of classical communication, it is easily seen that we can similarly use the state channel-duality, and more specifically the Gisin-Hughston-Jozsa-Wootters theorem \cite{GHJW1, GHJW2}, to provide the following alternative definition to definition~2.
\begin{def2b}
    
We say that the correlations $p(b|x,y)$ can be reproduced by an EA $d$-dimensional classical communication protocol if there exists 
	\begin{itemize}
		\item $d$ subnormalised states $\{\tau_{B}^{c,x}\}_c$ in $L(\mathcal{H}_B)$, where $B$ is a physical system with finite or separable Hilbert space $\mathcal{H}_B$, for each input $x$, where the total normalised state $\sum_{c=1}^d\tau_{B}^{c,x}=\tau_{B}$ is independent of $x$,
		\item an $n_{\text{B}}$-outcome POVM $\{M_{b|y,c}\}_b$ on $B$ for each input $y$ and $c\in [d]$,  
	\end{itemize}
	such that
	\begin{equation}\label{eq:corrclass2}
		p(b|x,y)=\sum_{c=1}^d \Tr\left(\tau_{B}^{c,x}\, M_{b|y,c}\right).
	\end{equation}
\end{def2b}

If we fix the dimension of $\mathcal{H}_B$ to some finite value $\text{dim}\left(\mathcal{H}_B\right)=D$, it is straightforward from definitions~1' and 2' that optimizing over the set of correlations $p(b|x,y)$ for fixed measurements is an SDP, as it amount to optimize over quantum states satisfying certain linear properties. Similarly, if we fix the states, the search for optimal measurements is also an SDP. Approximations to the set of correlations $p(b|x,y)$ can thus be obtained through a seesaw algorithm that repeatedly optimizes over the states for fixed measurements and then over the measurements for fixed states until some degree of convergence is achieved. This see-saw scheme represents an inner relaxation of the set of correlations for two reasons. First, though every solution that obtained is a valid strategy, it is not necessarily the optimal one. Second, some finite value $D$ on the dimension of $\mathcal{H}_B$ must be chosen. However, better solutions can in principle be obtained by increasing the dimension of $D$. If some general upper-bounds on the dimension of the shared entanglement were to hold and be known, this could evidently be used to limit the size of the SDP. 

More interestingly, we now provide SDP methods for obtaining outer relaxations that are valid irrespective of the amount of shared entanglement  i.e., without assumptions on $\text{dim}\left(\mathcal{H}_A\right)$ and $\text{dim}\left(\mathcal{H}_B\right)$. 

\subsection{Outer approximations through non-commutative polynomial optimization} \label{subsec-NPA}
To introduce our outer relaxation scheme, let us consider again definition~1. 
Since we do not assume any bound on the dimension of $\mathcal{H}_{A}$ or $\mathcal{H}_{B}$, the CPTP maps ${\$_{x} \colon L(\mathcal{H}_{A}) \to L(\mathcal{H}_{C})}$ appearing in Definition~1 can be seen as arising from unitary transformations on $\mathcal{H}_{A} \otimes \mathcal{H}_{C}$ and the POVMs $\{M_{b|y}\}_b$ can be assumed to be projective. That is, for some initial state $\ket{\varphi}_C$ in $\mathcal{H}_{C}$, we can re-express the correlations as
\begin{equation}
  \label{eq:pbxy-eaqc-clhs}
  p(b|x,y) = \bra{\Psi} U^{\dagger}_{x} M_{b|y} U_{x} \ket{\Psi} \,,
\end{equation}
where $\ket{\Psi} = \ket{\phi}_{AB} \ket{\varphi}_{C}$, the $U_{x} = U_{AC}^{x} \otimes \openone_{B}$ are unitaries that act nontrivially only on $\mathcal{H}_{A} \otimes \mathcal{H}_{C}$, and $M_{b|y} = \openone_{A} \otimes M_{BC}^{b|y}$ are projectors that act nontrivially only on $\mathcal{H}_{B} \otimes \mathcal{H}_{C}$.

Let us now introduce the following (Kraus) operators, which induce a parameterization on the system $C$,
\begin{IEEEeqnarray}{rCl}
  U_{A}^{x;j} &=& (\openone_A \otimes \bra{j}_C)U_x (\openone_A \otimes \ket{\varphi}_C) \,, \\
  M_{B}^{b|y;jk} &=& (\openone_B \otimes \bra{k}_C) M_{b|y} (\openone_B \otimes \ket{j}_C) \,.
\end{IEEEeqnarray}
Inserting two resolutions of the identity on $\mathcal{H}_C$ into \eqref{eq:pbxy-eaqc-clhs}, we find 
\begin{align}  
p(b|x,y) &= \sum_{j,k=0}^{d-1}\bra{\Psi} U_x^\dagger  \ketbra{j}{j} M_{b|y}  \ketbra{k}{k} U_x \ket{\Psi}  \\ 
  &= \sum_{j,k=0}^{d-1}\bra{\phi} U^{\dagger}_{x; j} U_{x; k}
  \otimes M_{b|y; jk} \ket{\phi} \,, \label{eq:probncp}
\end{align}
which now involves only subsystems $A$ and $B$, i.e., the shared state $\ket{\phi}$, the (Kraus) operators $U_{x;j}$ acting in $\mathcal{H}_{A}$, and the operators $M_{b|y; jk}$ acting in $\mathcal{H}_{B}$ (to simplify the notation, we drop the subsystem superscripts from the states and operators).

One can verify that the unitary conditions $U^{\dagger}_{x} U_{x} = \openone_{AC}$ translate to the operator constraints
\begin{IEEEeqnarray}{rCl}\label{eq:constrU}
  \sum_{k=0}^{d-1} U^{\dagger}_{x;k} U_{x;k} = \openone_A \,,
\end{IEEEeqnarray}
while the mathematical properties $\sum_{b} M_{b|y} = \openone_{BC}$ and $M^{\dagger}_{b|y} = M_{b|y}$, $M_{b|y} M_{b'|y} = \delta_{b,b'}M_{b|y}$ of the projectors are equivalent to
\begin{IEEEeqnarray}{rCl}
  \sum_{b} M_{b|y; jk} &=& \delta_{jk} \openone \,,\nonumber \\
  M^{\dagger}_{b|y; kj} &=& M_{b|y; jk} \,, \label{eq:constrM} \\
  \sum_{k=0}^{d-1} M_{b|y; jk} M_{b'|y; kl} &=& \delta_{bb'} M_{b|y; jl} \,.\nonumber
\end{IEEEeqnarray}

The problem of determining whether given correlations $p(b|x,y)$ can be reproduced through EA $d$-dimensional quantum communication, or finding the maximal value of a linear functional of the correlations $p(b|x,y)$, thus amounts to optimize over a state $\ket{\phi}$ and (non-Hermitian) operators $U_{x,j}$, $U^\dagger_{x,j}$, $M_{b|y;jk}$, satisfying the constraints \eqref{eq:constrU} and \eqref{eq:constrM} such that \eqref{eq:probncp} holds. Without the subsystem restriction and the tensor product appearing in \eqref{eq:probncp}, that would be a typical instance of noncommutative polynomial optimization \cite{NCpoly} to which the  Navascu\'es-Pironio-Ac\'in (NPA) hierarchy of SDP relaxations \cite{NPA2007,NPA2008} could be directly applied. As usual, one can relax the subsystem structure and the tensor product using instead commutation relations. That is, one can assume that all the operators $U_{x,j}$, $U^\dagger_{x,j}$, $M_{b|y;jk}$, act on the same Hilbert space, but satisfy
\begin{equation}
	[U_{x;j},M_{b|y;jk}]=0,\quad [U^\dagger_{x;j},M_{b|y;jk}]=0\,.
  \end{equation}
Physically, this amounts to considering a \emph{field-theoretic} variant of our prepare-and-measure scenario. The NPA hierarchy can now be applied directly. It provides an SDP relaxation hierarchy that represents an outer relaxation of the original, tensor-product variant of our problem, that converges asymptotically to the field-theoretic variant, and that returns the original tensor-product variant when rank optimality conditions are satisfied \cite{NCpoly,NPA2008}. 

Note that the scheme we just introduced can be seen as a hybrid scheme, where the dimension-free subsystems $A$ and $B$ are accounted for \emph{à la} NPA, while the subsystem $C$, whose dimension is fixed, is explicitly parameterized. In particular if we impose the additional constraints that all the operators on subsystems $A$ and $B$ commute between themselves (corresponding to a situation where the devices do not share any prior entanglement, but possibly only classical correlations), we recover a Lasserre-type SDP hierarchy \cite{Lasserre}  applied to an explicit parameterization of Alice's preparations of  and Bob's measurement on system $C$.

Finally, the case of EA classical communication can be seen as a sub-case of the general quantum communication by imposing additional constraints on Alice's operations forcing the output system $C$ to be in a diagonal state $\sum_{j=0}^{d-1} p(j|x)\ket{j}\bra{j}_C$ irrespectively of the input state $\ket{\phi}_{AB} \ket{\varphi}_{C}$. Alternatively and equivalently, one can directly start from definition~2. As we mentioned below that definition, the correlations in that scenario can be seen as the marginal correlations (obtained by summing over $c$) in a relaxed Bell scenario where the measurement performed on Bob's side depends not only on his input $y$ but also on the communicated output $c$ of Alice's measurement. Such relaxations of the usual Bell scenarios have been considered in \cite{chaves_unifying_2015,brask_bell_2017,himbeeck_quantum_2019}. Similarly to the observation made in \cite{himbeeck_quantum_2019} for the slightly different Instrumental scenario, one can then directly use the Navascu\'es-Pironio-Ac\'in (NPA) hierarchy for Bell nonlocal correlations \cite{NPA2007,NPA2008} in order to bound the correlations in our case. Indeed, it is immediate from \eqref{eq:corrclass} that the correlations $p(b|x,y)$ are a linear combination of standard Bell correlations where Bob has $n_\text{Y}\times d$ measurements labeled by inputs $y' = (y,c)$, and thus they can be viewed as linear combinations of entries of the moment matrices of the Bell-NPA hierarchy. 

\subsection{Outer approximations through information-restricted correlations}
The NPA relaxations that we have introduced above involve $2n_\mathrm{X}d+n_\mathrm{B}n_\mathrm{Y}d^2$ operators in the case of quantum communication and $n_\mathrm{X}d+n_\mathrm{B}n_\mathrm{Y}d$\footnote{This can be seen either directly from \eqref{eq:corrclass}, or by taking the classical limit of the quantum SDP introduced above (\ref{subsec-NPA}), where the assumption of classicality is encoded in the fact that Bob's POVM's $M_{b|y}$ are diagonal in a basis for $\mathcal{H}_C$.} operators in the case of classical communication. The size of the corresponding SDP is determined by the number of such operators and grows rapidly as one increases the order, with the number of such operators. In practice, these SDPs cannot be used to characterize EA communication scenarios with more than a few inputs or outputs without excessive computational resources.

For this reason, we propose an entirely different approach which applies equally well to both classical and quantum communication. It is based on two relaxations of the problem. Firstly, we relax the (post-communication) state space of Bob to a state space characterized by its information content \cite{InfoCorrelations}. Secondly, we use semidefinite relaxations of the set of informationally-restricted correlations \cite{InfoCorrelations2} to efficiently bound the correlations from EA communication. We now proceed to outline this approach.

Recently, a framework was developed for studying the correlations $p(b|x,y)=\Tr\left(\rho_x\,M_{b|y}\right)$ in prepare-and-measure experiments when the only assumption is that the \textit{information} relayed about $x$ through the states $\rho_x$ is restricted \cite{InfoCorrelations, InfoCorrelations2}. This information restriction is quantified through a bound on the guessing probability
\begin{equation}\label{pg}
	P_g=\max_{\{N_z\}}\sum_x p_x\Tr\left(\rho_x\,N_x\right),
\end{equation}
expressing how well the input $x$ can be guessed on average when performing an ideal measurement $\{N_z\}_z$ on the states $\rho_x$ if they are given with prior probabilities $p_x$. 
This restriction can equivalently be expressed in terms of the information quantity
\begin{equation}\label{info}
\mathcal{I}\equiv-\log_2\left(\max_x p_x\right)+\log_2 P_g,
\end{equation}
which quantifies in entropic terms the information that is gained when given the ensemble $\{p_x,\rho_x\}$ as compared to when the ensemble is not given (in which case the best guess of $x$ is its most likely value and this guess is thus correct with probability $\max_x p_x$). We refer the reader to Ref.~\cite{InfoCorrelations2} for further details on informationally-restricted correlations.

An important feature is that no assumption is made in \cite{InfoCorrelations2} on how the states $\rho_x$ that Bob eventually receives  and measures are physically prepared: they may leverage shared randomness, entanglement, high-dimensional systems etc. The only thing that matters is the bound on the information quantity \eqref{info}. Thus if we can find a bound on the information conveyed by the states $\rho_x=\tau_{{CB}}^x$ in \eqref{totalstate}, we can apply the methods of \cite{InfoCorrelations2} to our setting. 

In the case of classical communication, since the dimension of the message is $d$, we expect the bound $\mathcal{I}\leq \log d$ bits to be valid since, by no-signaling, shared entanglement should not help Alice communicate the value of $x$ to Bob. In the case of quantum dimension, we would instead expect a bound of  $\mathcal{I}\leq 2 \log d$ bits since shared entanglement can double the capacity of quantum communication through dense coding. Indeed, we now prove that this intuition is correct.

More generally, we will express a bound on $\mathcal{I}$ that depends on how well the entanglement shared between Alice and Bob is preserved by Alice's actions $\$_x$. For this let $k_x$ be the Schmidt number of the bipartite state $\tau_{{CB}}^x$ (i.e., the minimum Schmidt rank of the pure states in optimal ensemble realizing the density operator $\tau_{{CB}}^x$) and let $k=\max_x k_x$ be the largest Schmidt number. The parameter $k$ can be viewed as a measure of the entanglement content in the set of states $\tau_{{CB}}^x$. It satisfies the bounds $1\leq k\leq d$, where the lower limit corresponds to the case of classical communication, since the state $\tau_{{CB}}^x$ is a cq-state with no entanglement between system $C$ and system ${B}$, and the upper limit corresponds to the general case of quantum communication, where the bounds come from the fact that the local dimension of system $C$ is at most~$d$. 

\begin{proposition}
\label{prop:infobound}
	Consider a EA communication protocol where the total states $\tau_{{CB}}^x$ of Bob after Alice's communication are characterised by a maximal Schmidt number $k= \max_x k_x$. Then
	\begin{equation}\label{eq:infobound}
	\mathcal{I} \leq \log k +\log d\,,
	\end{equation}
	where $\mathcal{I}$ is the quantity defined in \eqref{info} and \eqref{pg}, and the above inequality is valid for any choice of the prior probabilities $p_x$. In particular, we get the bound $\mathcal{I}\leq \log d$ in the case of classical communication and the bound $\mathcal{I}\leq 2\log d$ in the case of general quantum communication. 
\end{proposition}
We note that a related, somewhat more restricted, result appears in the independent work~\cite{moreno}. We refer the reader to Appendix \ref{AppProp} for the proof.
    
Having introduced the connection to informationally-restricted correlations and their SDP relaxation hierarchy, several remarks are in order. First of all, though the relaxation to information-restricted correlations sometimes gives tight bounds (see examples in the next section), it represents in general a strict relaxation of the EA communication correlation set. Indeed, there exists correlations that can be obtained by Alice sending to Bob (high-dimensional) states that carry no more than $\mathcal{I}\leq 1$ bit of information, but which cannot be attained by the EA communication of a single bit (see example in the next section) \footnote{Similarly, we see no reason to expect that the relaxation to information would be tight in the case of EA \emph{quantum} communication. It is, however, much harder to find an explicit example, because of the high computational requirements of the NPA-type hierarchy for EA quantum communication.}. The approach based on the NPA hierarchy inherits, in contrast, its nice converging properties and we thus expect that it will in general give better bounds at a sufficiently high order. However, in practice, when taking into account limited computer memory and time, the information-based SDP relaxation may sometimes be superior. The reason is that, as we pointed out earlier, the size of these SDP relaxations grows rapidly with the number of basic operators involved, which depend on a factor of order $d^2$ in the case of quantum communication and $d$ in the case of classical communication.  In contrast, the information-based SDP relaxation have a much smaller size, which is moreover independent of $d$. This provides an advantage for the later relaxation both for fixed dimension $d$ when one increase the relaxation order and at fixed relaxation order when one increases the dimension $d$. 

Finally, note that in a situation where no entanglement is pre-shared between Alice and Bob the Schmidt number $k$ satisfies $k=1$ even in the case of quantum communication. Thus the bound $\mathcal{I}\leq \log d$ valid for EA $d$-dimensional classical communication is also valid for non-EA $d$-dimensional quantum communication. Thus the SDP relaxation hierarchy based on  informationally-restricted correlations does not distinguish these two sets of correlations. We will come back to the relationship between these two sets in section~\ref{sec:resources} and see that in general they are distinct, overlapping sets (in particular, the set of EA classical $d$-dimensional correlations is not contained in the set of non-EA quantum $d$-dimensional correlations and vice-versa). 
Similarly, the bound $\mathcal{I}\leq 2\log d$ valid for EA $d$-dimensional quantum communication is also valid for non-EA $d^2$-dimensional quantum communication and thus the SDP relaxation hierarchy based on  informationally-restricted correlations does not distinguish these two sets. Again we discuss in more detail the relationship between these two sets in section~\ref{sec:resources}.

\section{Application: revised classical and quantum tests of dimension}\label{sec:applications}
We now apply the methods introduced in the previous section towards the task of device-independently testing the dimension of a physical system (classical or quantum). Specifically, we consider two different tests of dimension that have been previously investigated, in both theory and experiment, and re-examine their analysis to account for the most general picture in which parties may share unlimited entanglement. Notably, in all cases we consider, we obtain either optimal or close to optimal dimension witnesses.

\subsection{The Random Access Code}
Let us begin with the dimension witness  experimentally realised in Ref~\cite{Ahrens2014}. It is based on the regular Random Access Code \cite{Ambainis1999} introduced at the beginning of section~\ref{sec:flag} in which Alice has a choice among four possible inputs $x\in[4]$, Bob has a binary input $y\in[2]$ and generates binary outcomes $b\in[2]$. We are interested in the success probability of this RAC, which can be expressed through the RAC correlation function $W_\text{RAC}$ defined in eqs.~\eqref{Wexp} and \eqref{coefs}.
When no entanglement is present, the following bounds on $W_\text{RAC}$ for classical and quantum systems of dimension two, three, and four are known \cite{Ahrens2014}:
\begin{equation}\label{oldRACbound}
W_\text{RAC}\hspace{1mm}\stackrel{\text{C2}}{\leq}\hspace{1mm} 4 \hspace{1mm} \stackrel{\text{Q2}}{\leq}\hspace{1mm} 4\sqrt{2}\hspace{1mm} \stackrel{\text{C3}}{\leq}\hspace{1mm} 6 \hspace{1mm}\stackrel{\text{Q3}}{\leq}\hspace{1mm} 2\left(1+\sqrt{5}\right)\hspace{1mm}\stackrel{\text{C4, Q4}}{\leq}\hspace{1mm} 8.        
\end{equation}
Notice that for four-dimensional classical and quantum systems, Alice may simply send her input to Bob and thus one reaches the algebraically maximal value of $W_\text{RAC}=8$. Depending on which inequalities in the chain \eqref{oldRACbound} are experimentally violated, one can certify that systems of certain minimal classical and quantum dimensions have been produced. 

Let us re-examine the problem in a fully device-independent setting in which Alice's and Bob's device may share prior entanglement. First, it is clear that quantum dense coding allows Alice to send her entire input to Bob and thus reach the algebraically maximal value of $W_\text{RAC}=8$ using only EA qubit communication, i.e. the following bound is tight:
\begin{equation}
	W_\text{RAC}\hspace{1mm}\stackrel{\text{Ent-Q2}}{\leq}\hspace{1mm} 8\,.
\end{equation}
Therefore, when entanglement is allowed, it is no longer possible to certify three- and four-dimensional quantum communication using the RAC dimension witness $W_\text{RAC}$.

In the classical case, the re-examination is less trivial. In order to bound $W_\text{RAC}$ for EA classical communication of dimensions two and three, we have evaluated semidefinite relaxations based on both the NPA hierarchy\footnote{For $d=2$ we use level $1$ and for $d=3$ we use level $1+AB+AA$.}
 and the information-based SDP hierarchy\footnote{For $d=2$ the moment matrix was size $39$ and the localising matrix was size $8$. For $d=3$, the sizes were $87$ and $20$ respectively.}
  for both messages of dimension $d=2$ and $d=3$. We obtain the same bounds with both approaches, namely
\begin{equation}\label{RACbound}
W_\text{RAC}\hspace{1mm}\stackrel{\text{Ent-C2}}{\leq}\hspace{1mm}5.657 \hspace{1mm} \stackrel{\text{Ent-C3}}{\leq} \hspace{1mm}6.828.
\end{equation}
The first bound is tight (up to solver precision), since it is known that it can be saturated if Alice and Bob use the shared entanglement to maximally violate the Clauser-Horne-Shimony-Holt Bell inequality before communicating a two-dimensional classical system \cite{Buhrman2001}. Moreover, we find that also the second bound is tight. To show this, we have used the SDP seesaw routine described in the previous section to efficiently search for optimal EA classical communication strategies. This leads us to find an explicit strategy, involving a shared entangled state of local dimension $D=4$, achieving $W_\text{RAC}=6.828$. These results  show that in order to test classical dimension in the presence of entanglement, one must significantly revise the bounds in Eq.~\eqref{oldRACbound} which are valid only when entanglement is assumed to not be present in the experiment.

In order to also consider a non-trivial setting for EA quantum communication, let us return to the modified RAC considered in section~\ref{sec:flag}, where Alice and Bob each are supplied with one more input and asked to maximize \eqref{Wexp} under the constraint \eqref{constraint}. Rather than requiring these constraints to be exactly satisfied, we can instead incorporate them in a modified witness
\begin{equation}\label{Wexp2}
		W_\text{fRAC}=\sum_{x=1}^5\sum_{y=1}^3 c_{xy}E_{xy},
		\end{equation}
		where the $5\times 3$ coefficients $c_{xy}$ are given by 
		\begin{align}\label{coefs2}
		& c=\begin{pmatrix}
		1 & 1 & \beta\\
		1 & -1 & \beta \\
		-1 & 1 & \beta\\
		-1 & -1 & \beta\\
		0 & 0 & -4\beta
		\end{pmatrix} \,,
		\end{align}
depending on some positive parameter $\beta$, which favours (when $\beta$ is sufficiently large) strategies where the constraints \eqref{constraint} are satisfied. For concreteness, we used $\beta=4$. The explicit qubit strategy using entanglement of local dimension $4$ that we introduced in section~\ref{sec:fourdim} achieves a value $W_\text{fRAC}=6.828+8\beta=38.8284$. 

We evaluated the information-based SDP relaxation\footnote{Moment matrix size $866$. Localising moment matrix size $127$.} for EA qubit communication and obtained 
\begin{equation}
W_\text{fRAC}\hspace{1mm}\stackrel{\text{Ent-Q2}}{\leq}\hspace{1mm}38.8284.
\end{equation} 
This result is (up to solver precision) identical to that obtained using the explicit strategy of section~\ref{sec:fourdim}, showing that it is optimal. On the other hand, an EA qutrit (allowing to communicates two classical trits using dense coding) can reach the algebraic maximum $W_\text{fRAC}=8+8\beta=40$. The modified witness \eqref{Wexp2} thus constitutes a proper qutrit witness, even in the presence of arbitrary entanglement.

\subsection{The Witness of Gallego \textit{et al.}}\label{sec:gallego}
Let us now consider another dimension witness, introduced in \cite{Gallego2010}, different variants of which have been experimentally realized in \cite{Ahrens2012, Hendrych2012, Ambrosio2014}. 

In this scenario, Alice receives one of five possible inputs $x\in[5]$, Bob receives one of four possible inputs $y\in[4]$ and produces binary outcomes $b\in[2]$. The witness, labelled $W_\text{5}$, is written in the correlation format of \eqref{Wexp} with coefficients
\begin{equation}
c=\begin{pmatrix}
1 &   1  &   1   &  1\\
1  &   1 &    1 &   -1\\
1  &   1   & -1  &   0\\
1  &  -1  &   0  &   0\\
-1   &  0  &   0 &    0
\end{pmatrix}.
\end{equation}
In a scenario without shared entanglement, the tight bounds on the witness for classical systems were  obtained in \cite{Gallego2010}:
\begin{equation}\label{GallegoboundsC}
W_\text{5}\hspace{2mm}\stackrel{\text{C2}}{\leq}\hspace{2mm} 8\hspace{2mm} \stackrel{\text{C3}}{\leq}\hspace{2mm} 10\hspace{2mm} \stackrel{\text{C4}}{\leq} \hspace{2mm}12\hspace{2mm}\stackrel{\text{C5}}{\leq}\hspace{2mm} 14.   
\end{equation}
In addition, using symmetrised semidefinite relaxations \cite{TavakoliSymmetry}, we have computed upper bounds on $W_\text{5}$ for dimensionally restricted quantum systems without shared entanglement. These bounds are tight, since we could  saturate them numerically with explicit quantum strategies.
\begin{equation}\label{GallegoboundsQ}
W_\text{5}\hspace{1.5mm}\stackrel{\text{Q2}}{\leq}\hspace{1.5mm} 8.828\hspace{1.5mm} \stackrel{\text{Q3}}{\leq}\hspace{1.5mm} 11.527\hspace{1.5mm} \stackrel{\text{Q4}}{\leq} \hspace{1.5mm}13.036\hspace{1.5mm}\stackrel{\text{Q5}}{\leq}\hspace{1.5mm} 14.   
\end{equation}
Hence, the witness $W_\text{5}$ enables certifying systems of classical and quantum dimension of two, three, four or five in scenarios without shared entanglement.

Let us re-consider this analysis for the situation in which entanglement may be shared between Alice and Bob. We first note that in the quantum case the maximal algebraic value $W_\text{5}=14$ can be attained with EA qutrit communication since a dense coding protocol can be used to relay $x$ to Bob, i.e., the following tight bound holds:
\begin{equation}
	W_\text{5}\hspace{1.5mm}\stackrel{\text{Ent-Q3}}{\leq}\hspace{1.5mm} 14\,.
\end{equation}

For classical communication of dimension two, three and four as well as for quantum communication of dimension two, we reanalyze the classical bounds \eqref{GallegoboundsC} and the qubit bound \eqref{GallegoboundsQ} in the presence of shared entanglement.

We start by considering explicit classical and quantum communication strategies based on sharing  two copies of the maximally entangled two-qubit state. Using the seesaw routine of the previous section, we have found explicit strategies achieving $W_\text{5}=9.034$ (Ent-C2), $W_\text{5}=11.515$ (Ent-C3) and $W_\text{5}=13.036$ (both Ent-C4 and Ent-Q2), violating the non-entanglement-assisted bounds for all considered resources. 

Next, we computed upper bounds on $W_5$ for these entanglement-assisted resources using our new SPD methods. In the classical case, we have again considered both the NPA hierarchy and the information-based SDP hierarchy. This time, however, the two methods return different results. The best bounds we obtained are the following:
\begin{equation}\label{GallegoboundsNew}
W_\text{5}\hspace{1mm}\stackrel{\text{Ent-C2}}{\leq}\hspace{1mm} 9.034^{{\color{blue}\text{npa}}}\hspace{1mm} \stackrel{\text{Ent-C3}}{\leq}\hspace{1mm} 11.563^{{\color{blue}\text{info}}}  \stackrel{\text{ Ent-C4}}{\leq} 13.095^{{\color{blue}\text{info}}} ,   
\end{equation}
where ${{\color{blue}\text{npa}}}$ indicates that the result was obtained with the NPA SDP relaxation\footnote{With NPA, we obtain $9.034$, $11.568$ and $13.225$ respectively. For $d=2,3$ we use the second level. For $d=4$ we restrict to level $1+AB$ due to the increasingly demanding computation.} and ${{\color{blue}\text{info}}}$ indicates that the result was obtained with the information-based SDP relaxation\footnote{With information relaxations, we obtain $9.081$, $11.563$ and $13.095$ respectively, using moment matrices and localizing matrices of size $498$ and $91$ respectively.}. Thus the lower and upper bounds nearly match: their ratios are  $>99.9\%$, $99.5\%$, $99.5\%$ and $99.5\%$ respectively.  Thus, we find the NPA SDP relaxation performs better for the smaller problem ($d=2$) but that the information-based SDP relaxation becomes advantageous for the somewhat larger problem ($d=3,4$). Furthermore, the bound for $d=2$ allows us to prove that the information relaxation of EA classical communication is, generally, not tight. Indeed, by numerical search we have found explicit strategies from quantum states carrying at most $\mathcal{I}=1$ bit of information, that achieves $W_\text{5}=9.054$. Thus the information-based SDP relaxation, even if implemented at an arbitrary high order, cannot return an upper-bound smaller than $W_\text{5}=9.054$. But this exceeds the upper-bound \eqref{GallegoboundsNew} on EA classical communication for $d=2$ found using NPA. 

For the case of EA qubit communication, the relevant information-based SDP relaxation is identical to that considered for EA quart communication. Therefore, the following quantum dimension witness is immediately obtained 
\begin{equation}\label{Gallegobounds2}
W_\text{5}\stackrel{\text{Ent-Q2}}{\leq} 13.095.   
\end{equation}
A quantum system of minimal dimension three can be certified by violating this inequality. For comparison, solving the NPA relaxation for EA quantum communication for the same problem at level 2 of the hierarchy we recovered only the trivial algebraic bound $W_\text{5} \lesssim 14$ after more than 40 hours of computing time.

To summarize, we emphasize that the bounds \eqref{GallegoboundsC} and \eqref{GallegoboundsQ} are completely revised, both in the classical and quantum case,in a fully DI setting where entanglement is permitted in the experiment.

\section{Quantum communication versus EA classical communication}\label{sec:resources}
A (non-EA) qubit and a EA bit represent two different ways of employing quantum resources to generate correlations in a prepare-and-measure setting, which in both cases cannot be used to communicate more than one bit of information on Alice's input (in particular the use of the information-based SDP hierarchy leads to the same relaxations of both sets). It is natural to ask how these resources compare, and in particular if one is more powerful than the other.

\subsection{EA bits can outperform qubits}
The comparison of qubits and EA classical bits has already received substantial research attention. It was proven in \cite{Pawlowski2012} that when Bob has binary outcomes, all correlations obtained by communicating a qubit can be simulated by communicating an EA bit\footnote{It is further claimed in \cite{Pawlowski2012} that EA bit strategies are equivalent to \emph{hyperbit} strategies. However, while it is true that EA bit strategies can simulate any hyperbit strategy, the converse is wrong. Using the techniques introduced in the present manuscript, we found explicit EA bit correlations that cannot be simulated by hyperbit strategies. See Appendix \ref{Apphyperbits} for details.}.

By combining the results of \cite{Ambainis2006} and \cite{Pawlowski2010} it follows that this resource inequality is strict. Ref~\cite{Pawlowski2010} shows that EA classical bits achieve a winning probability of $p_\text{win}=3/4$  four-bit RAC with binary communication, while \cite{Ambainis2006} shows that  non-EA qubits must satisfy $p_\text{win}<3/4$. However, the noise-tolerance of this advantage is presently restricted only to numerical evidence \cite{Ambainis2006}. 

Our results from section~\ref{sec:gallego} in fact prove a noise-robust gap between the correlations obtained from non-EA qubits and EA classical bits. Specifically, in the former case we found that the witness obeys $W_\text{5}\leq 8.828$ while in the latter case it can achieve $W_\text{5}=9.034$. This enables experimental certification of the advantage of EA classical communication which tolerates substantial amount of noise.

\subsection{Qudits can outperform EA classical dits}
A more interesting situation is encountered when one goes beyond binary outcomes for Bob. Several different works  \cite{Tavakoli2016, Hameedi2017, Tavakoli2017b, Martinez2018} hint that the above situation can be reversed, i.e.~that non-EA qudit communication can outperform EA classical dit communication. However, all these works consider a certain  subclass of EA classical dit strategies, which may not always do justice to the full power of EA classical communication \cite{Brukner2020}. We now proceed to employ the general tools developed in section~\ref{sec:methods} to prove that non-EA qudit communication can outperform completely general EA classical dit communication.

We consider the scenario of \cite{Tavakoli2016}, which is a higher-dimensional version of the previously considered RAC. In this task, Alice has nine possible inputs $x\in[9]$ represented by two trits $x_1,x_2\in\{0,1,2\}$. Bob has an input $y\in[2]$ and aims to guess Alice's $y$'th input trit in his output $b\in[3]$. With uniformly distributed inputs, the average success probability is 
\begin{equation}\label{RACdim}
W=\frac{1}{18}\sum_{x_1,x_2=0}^{2}\sum_{y=1}^{2}p(b=x_y|x_1,x_2,y).
\end{equation}
It is shown in \cite{Tavakoli2015} that non-EA qutrit communication can achieve $W=\frac{1}{2}\left(1+\frac{1}{\sqrt{3}}\right)\approx 0.788$ and in \cite{Tavakoli2016} that certain class of EA trit communication protocols based on a natural Bell inequality violation satisfy the bound $W\leq \frac{7}{9}\approx 0.778$. 

We now employ our general methods to re-examine the analysis of EA trit communication without the additional assumption of restricting to a particular Bell inequality test. First, we use the seesaw routine described in section~\ref{sec:methods}. Considering entanglement of local dimension $D=3$, we recover the value $W=\frac{7}{9}$. However, by considering entanglement of local dimension $D=9$, we are able to find an improved protocol that achieves $W=0.784$. This shows that the EA classical communication protocol considered in \cite{Tavakoli2016} are not general enough.

This leads to the question of whether non-EA qutrits do actually outperform the most general EA trit protocols. We answer this in the positive by employing our information-based SDP relaxations to bound $W$ for the latter case. To enable this computation on a standard desktop computer, we have exploited the symmetries of the function \eqref{RACdim} to reduce the number of variables in the final SDP matrix\footnote{The moment matrix is size $1585$ and originally featured over $5\times 10^5$ variables. The use of symmetries reduced it to only about $5\times 10^3$ variables.}. We find the bound
\begin{equation}
W \stackrel{\text{EA-C3}}{\leq} 0.787.
\end{equation}
This upper bound is likely to be only nearly optimal, but even so it is strictly smaller than the qutrit protocol achieving $W=0.788$. Thus, it proves that non-EA quantum communication can outperform fully general EA classical communication.

\section{Conclusions}
In this work, we have investigated the correlations that can be generated in prepare-and-measure experiments in which parties share entanglement and communicate either classical or quantum systems of a given dimension. We showed that the strongest forms of quantum correlations require protocols that go beyond the paradigmatic quantum dense coding protocol and developed general methods for bounding the correlations that can be obtained in such experiment when an unlimited amount of entanglement is allowed. We applied this to introduce device-independent tests of the dimension of classical and quantum systems that make no assumption on the presence of entanglement between the involved devices and showed how this warrants a re-examination of standard tests of dimension in which entanglement is assumed not be present in experiments. We also applied the methods to investigate the relation between entanglement-assisted communication protocols and non-entanglement-assisted communication protocols.

Our work introduces the main conceptual and technical tools necessary to pave the way for a systematic investigation of entanglement-assisted communication in prepare-and-measure scenarios. It  leaves open several natural questions and continuations. First, while the method for bounding correlations based on a relaxation to informationally restricted quantum correlations is relatively efficient computationally and, as we have seen, often leads to strong bounds, one cannot in general expect the bounds to be optimal. How can one overcome this limitation in both a conceptual and practical manner? Second, we have shown that the strongest correlations possible from EA qubit communication in general require high-dimensional entanglement\footnote{This is true even if shared randomness is considered a free resource as we proved this result using a linear dimension witness for which extremal strategies are always optimal}. How high does this dimension need to be? Does there exist correlations that can only be generated with qubit communication and infinite-dimensional entanglement? Third, our work motivates a comparison of different type of quantum resources in prepare-and-measure scenarios. We have explored some of these in section~\ref{sec:resources}. But there remain many other open questions. For instance, our results in section ~\ref{sec:beyonddc} imply that there exists scenarios where EA qubit communication can outperform non-EA quaquart communication (compare eq. \eqref{res2} to \eqref{res})\footnote{This is in contrast to the case of classical communication, where it was shown that non-EA quarts (with shared randomness) can always be used to simulate bits assisted by any shared maximally entangled state \cite{frenkel2021}.}. Is this a strict resource inequality or does there exists scenarios in which non-EA ququart communication can outperform EA qubit communication. Fourth, it is interesting to explore tests of classical and quantum dimension when entanglement is involved. Can one construct simple families of dimension witnesses, favorably based on binary measurements, that are valid for any dimension? Fifth, tests of physical dimension are primarily practically motivated tasks. This has lead to several experiments (see e.g.~\cite{Ahrens2012, Hendrych2012, Ahrens2014, Ambrosio2014}) testing dimension in standard prepare-and-measure experiments (assuming no shared entanglement). However, as we have shown, the conclusions of such experiments are generally not valid when entanglement is introduced. It is interesting and relevant to consider experimental implementations of device-independent tests of both classical and quantum dimensions when no assumptions are made on the entanglement that can be shared between preparation and measurement devices. 
Finally, it would also be interesting to use our method to design and analyze the security of semi-device-independent quantum random number generation and quantum key distribution protocols that do not make the assumption that the devices do not share prior entanglement.

\section*{Acknowledgements}

This work was supported by the Swiss National Science Foundation through Early PostDoc Mobility fellowship P2GEP2 194800, the EU Quantum Flagship
project QRANGE and the Fonds National de la Recherche Scientifique F.R.S.-FNRS (Belgium) under a FRIA grant. S.P. is a Senior Research Associate of
the Fonds de la Recherche Scientifique - FNRS.

\newpage
\appendix
\section{Proof of Proposition~\ref{prop:infobound}}\label{AppProp}

 To prove Proposition~\ref{prop:infobound}, we need the following Lemma.
\begin{lemma1}
	Let $\rho$, $\sigma$ be two positive semidefinite operators on a joint, finite Hilbert space $\mathcal{H}_C \otimes \mathcal{H}_{B}$.  Then
	\begin{equation} 
		\Tr[\rho \sigma] \leq k \Tr [\rho_{B}\sigma_{B}] \label{eq:lemma1},
		\end{equation}
		where $k$ is the Schmidt number of $\rho$ and $X_{B} = \Tr_C[X]$ denotes the partial trace. 
\end{lemma1}
\begin{proof}
	Let us start by assuming that $\rho=\ketbra{\phi}{\phi}$ and $\sigma=\ketbra{\psi}{\psi}$ are rank one. We can then introduce Schmidt decompositions
	\begin{align}
	\ket{\phi} &= \sum_{i=1}^k \sqrt{p_i} \ket{\alpha_i}\ket{\beta_i}, \\
	\ket{\psi} &= \sum_j \sqrt{q_j} \ket{\alpha_j'}\ket{\beta_j'}.
	\end{align}
	where the sum over $i$ runs at most over $k$ values by assumption. 
	
	We have that $\Tr[\phi\psi] = \abs{\braket{\phi}{\psi}}^2$.
	Explicitly writing the inner product gives
	\begin{align}
	\abs{\braket{\phi}{\psi}} &= \Babs{\sum_{ij} \sqrt{p_i}\sqrt{q_j} \braket{\alpha_i}{\alpha_j'}\braket{\beta_i}{\beta_j'}} \nonumber \\ &\leq \sqrt{\sum_{ij}\abs{\braket{\alpha_i}{\alpha_j'}}^2}\sqrt{\sum_{ij} p_iq_j\abs{\braket{\beta_i}{\beta_j'}}^2} \nonumber \\ &\leq  \sqrt{k}\sqrt{\Tr[\phi_{B}\psi_{B}]}.
	\end{align}
	The second line follows from applying the Cauchy-Schwarz inequality $\abs{\mathbf{a} \cdot \mathbf{b}} \leq \norm{\mathbf{a}} \norm{\mathbf{b}}$ to vectors of components 
	\begin{align}
	\mathbf{a} &= \bigl( \braket{\alpha_i}{\alpha_j'}^{*} \bigr), \\
	\mathbf{b} &= \bigl( \sqrt{p_i}\sqrt{q_j} \braket{\beta_i}{\beta_j'} \bigr).
	\end{align}
	The third line follows by noting that $\sum_{ij}\abs{\braket{\alpha_i}{\alpha_j'}}^2\leq k$ since the $\ket{\alpha'_j}$ form an orthornomal basis and the sum over $i$ runs over at most $k$ values. Equally is thus obtained when $i$ runs over precisely $k$ values.
	Hence, \begin{equation}\label{eq:rank1}
	\Tr[\phi\psi] \leq k \Tr [\phi_{B}\psi_{B}]. 
	\end{equation}
	If $\rho$ and $\sigma$ are not rank one, we can decompose them as $\rho=\sum_i\phi_i$ and $\sigma = \sum_j \psi_j$ where the $\phi_i$ and $\psi_j$ are rank one and furthermore the Schmidt rank of $\phi_i$ is at most $k$ by assumption. Then using that the relation \eqref{eq:rank1} is linear, we get
	\begin{align}
	\Tr [ \rho \sigma]  &= \sum_{ij} \Tr[\phi_i \psi_j] \nonumber \\ &\leq k \sum_{ij} \Tr[\phi_{i{B}} \psi_{j{B}}] \nonumber \\ &=k \Tr[\rho_{B} \sigma_{B}]\,.
	\end{align}
\end{proof}

Equipped with this, we can now prove a general bound on the guessing probability \eqref{pg} for the specific states $\rho_x=\tau_{{CB}}^x$ in our EA scenario:
\begin{align}\nonumber
P_g&=\max_{\{N_z\}}\sum_x p_x\Tr\left(\tau_{{CB}}^x\,N_x\right)\\\nonumber
&\leq k\max_{\{N_z\}}\sum_x p_x \Tr\left(\tau_{B}\, \tilde N^x\right)\\\nonumber
&\leq k\left(\max_xp_x\right)\max_{\{N_z\}} \Tr\left(\tau_{B}\, \sum_x \tilde N_x\right)\\\nonumber
& =k\left(\max_xp_x\right)\max_{\{N_z\}} \Tr\left(\tau_{B}\, \Tr_C\left(\openone_C\otimes\openone_B\right)\right)\\
& =kd\left(\max_xp_x\right). \label{eq:Qinfobound}
\end{align}
In the second line we used the inequality \eqref{eq:lemma1}, together with the fact that $\tau_{B}^x=\tau_{B}$ is independent of $x$, and introduced the notation $\tilde N_x=\Tr_C(N_x)$. In the fourth line we have used the completeness of the measurement $\{N_z\}$ and in the fifth line the fact that $\Tr_C \openone_C =d$ since $\text{dim}(\mathcal{H}_C)=d$. Inserting this bound into \eqref{info}, we get \eqref{eq:infobound}.

We have thus relaxed the problem of deciding whether $p(b|x,y)$ can be achieved by EA communication of a classical or quantum $d$-dimensional system to a problem of deciding whether the same correlations can be achieved by states carrying at most, respectively, $\log d$ or $2\log d$ bits of information. The latter problem is known to admit a hierarchy of increasingly precise semidefinite relaxations \cite{InfoCorrelations2}, which can thus also be applied to our original problem. In Appendix.~\ref{AppSDP}, we discuss the main features of this hierarchy. 

The information bound \eqref{eq:infobound} is valid for any choice of the prior probabilities $\{p_x\}$ and any such choice defines a relaxation of the set of correlations achievable through EA $d$-dimensional communication. However, the choice of uniform priors,  i.e., $p_x=\frac{1}{n_\text{X}}$, is the optimal choice that results in the most constraining relaxation. Indeed, any bound of the form $\mathcal{I}\leq \alpha$ for arbitrary priors is necessarily implied by the bound $\mathcal{I}_\text{uni}\leq \alpha$ for uniform priors $p_x=\frac{1}{n_\text{X}}$. To see this, simply note that 
\begin{align}
	P_g^\text{bias}&=\max_{\{N_z\}} \sum_{x=1}^{n_\text{X}} p_x\Tr\left(\rho_xN_x\right) \nonumber\\
	&\leq \left(\max_x p_x\right)\max_{\{N_z\}} \sum_{x=1}^{n_\text{X}} \Tr\left(\rho_xN_x\right) \nonumber \\
	& = \left(\max_x p_x\right) n_\text{X} P_g^\text{uni}.
	\end{align}
	Consequently,
	\begin{align}
	\mathcal{I}&\leq -\log \left(\max_x p_x\right)+\log\left(\left(\max_x p_x\right) n_\text{X}P_g^\text{uni}\right) \nonumber\\
	&=\log n_\text{X}+\log\left(P_g^\text{uni}\right)=\mathcal{I}_\text{uni}\leq \alpha.
	\end{align}

\section{Sketch of the information-based SDP hierarchy}\label{AppSDP}
Here, we sketch the  hierarchy of semidefinite programs for bounding the set of informationally-restricted quantum correlations, that we have employed to efficiently (but in general not tightly) bound the set of correlations obtainable from classical or quantum communication in entanglement-assisted prepare-and-measure scenarios. This hierarchy is originally introduced in Ref.~\cite{InfoCorrelations2} to investigate the concept of informationally-restricted correlations. Notably, due to a connection between this concept and quantum contextuality, a modification of this hierarchy was also recently proposed  to bound the set of quantum correlations in contextuality experiments \cite{Abbott2021}.

Given that Alice's input is sampled from a probability distrubution $p_x$, the information carried by her communicated quantum ensemble is given by Eq.~\eqref{info}. This quantity is one-to-one with the so-called guessing probability $P_g$, defined in Eq.~\eqref{pg} as the performance of the best possible quantum protocol for minimal-error state discrimination of Alice's ensemble.  Consider now that we are given a probability distribution $p(b|x,y)$ in a prepare-and-measure scenario and asked to decide whether $p(b|x,y)$ is compatible with some quantum model based on an information transmission corresponding to a fixed value of $P_g$ (for given $p_x$). Ref.~\cite{InfoCorrelations2} presented a hierarchy of increasingly precise necessary conditions for the existence of such a quantum model. Each condition takes the form of a semidefinite program. We now proceed to sketch the construction of these semidefinite relaxations.

Define a list of operators containing the all preparations and measurements in the relevant scenario: 
\begin{equation}
S=\{\openone, \sigma, \rho_1,\ldots,\rho_{n_\text{X}},M_{1|1},\ldots,M_{n_\text{B}|1},\ldots,M_{n_\text{B}|n_\text{Y}}\}.
\end{equation}
The measurements can without loss of generality be chosen as projective ($M_{b|y}M_{b'|y}=\delta_{b,b'}M_{b|y}$). Moreover, the list also includes the identity (of unknown dimension) and an auxiliary operator $\sigma$. By considering products of the elements of $S$ (monomials), we can build a list, which we name $\mathcal{S}$. This list should, at the very least, contain all elements of $S$ (products of length one). Adding more monomials to this list will correspond to a more precise necessary condition for a quantum model. From the momomial list $\mathcal{S}$, we can now build a $|\mathcal{S}|\times |\mathcal{S}|$ matrix of moments, defined as
\begin{equation}
\Gamma_{ij}=\Tr\left(\mathcal{S}_i\mathcal{S}_j^\dagger\right).
\end{equation}
Note that the quantum probabilities $p(b|x,y)=\Tr\left(\rho_xM_{b|y}\right)$ appear as explicit entries in the moment matrix. By associating $\Tr\left(\rho_xM_{b|y}\right)$ to the corresponding entry $(i,j)$ in $\Gamma$, we may label the relevant entries as  $\Gamma_{bxy}$. A quantum model is compatible with the positivity condition $\Gamma\geq 0$.

So far, no physically relevant constraint has been placed on $\Gamma$. A priori, it may appear difficult to impose the constraint on the guessing probability since it itself corresponds to a semidefinite program. The key observation for resolving the apparent difficulty is that one may exploit the program dual to that corresponding to the guessing probability. Specifically, define $\sigma\geq p_x \rho_x$ $\forall x$. Then,
\begin{equation}
P_g=\max_{\{N_z\}}\sum_x p_x \Tr\left(\rho_x N_x\right) \leq \max_{\{N_z\}}\sum_x \Tr\left(\sigma N_x\right)=\Tr\left(\sigma\right).
\end{equation}
This is the reason why we have included $\sigma$ in the operator list $S$. In order to constrain the guessing probability, we can impose a bound on $\Tr\left(\sigma\right)$, which appears as an explicit entry in $\Gamma$. We label that single entry as $\Gamma_\sigma$. However, in order to nontrivially impose this constraint, we must also account for the semidefinite conditions  $\sigma\geq p_x \rho_x$ on the level of the semidefinite relaxation. To this end, we introduce a set of localising matrices defined as
\begin{equation}
\tilde{\Gamma}^{(x)}_{ij}=\Tr\left(\mathcal{R}_i\left(\sigma-p_x\rho_x\right)\mathcal{R}_j^\dagger\right),
\end{equation}
where $\mathcal{R}$ is (in analogy with $\mathcal{S}$) a list of monomials built from products taken from the elements of $S$. Notably, one does not need to choose $\mathcal{R}=\mathcal{S}$ but it is favourable to choose $\mathcal{R}$ such that all entries in $\tilde{\Gamma}^{(x)}$ appear in $\Gamma$. Imposing the positivity constraint $\tilde{\Gamma}^{(x)}\geq 0$ for all $x$ nontrivially enforces the desired constraints.

Finally, it is known that the analysis of informationally-restricted correlations cannot be restricted to pure quantum states without loss of generality \cite{InfoCorrelations2}. This means that we do not wish to enforce $\rho_x^2=\rho_x$ but instead the constraint $\rho_x\geq 0$ on the level of the semidefinite relaxation. To this end, we introduce another set of localising matrices, defined as
\begin{equation}
\bar{\Gamma}^{(x)}_{ij}=\Tr\left(\mathcal{T}_i \rho_x\mathcal{T}_j^\dagger\right),
\end{equation}
where $\mathcal{T}$ is (in analogy with $\mathcal{S}$ and $\mathcal{R}$) a list of monomials built from products taken from the elements of $S$. The positivity of the state is nontrivially imposed by the condition $\bar{\Gamma}^{(x)}\geq0$.

Putting the above togther, a necessary condition for the existence of a quantum model with guessing probability $P_g$ (for given prior $p_x$) takes the form of the following semidefinite program:
\begin{IEEEeqnarray}{l+l}
\text{Find } \{\Gamma,\tilde{\Gamma}^{(x)},\bar{\Gamma}^{(x)}\} \text{ such that }\nonumber \\
\qquad \Gamma\geq 0,\quad  \forall x:\hspace{1mm} \tilde{\Gamma}^{(x)}\geq 0,\quad  \forall x: \hspace{1mm} \bar{\Gamma}^{(x)}\geq 0 \nonumber\\
\qquad \Tr\left(\rho_x\right) = 1, \quad \Gamma_\sigma\leq P_g, \quad \Gamma_{bxy}=p(b|x,y) \label{eq:poloptim2}.
\end{IEEEeqnarray}
For any given choice of monomial lists $\{\mathcal{S},\mathcal{R},\mathcal{T}\}$, the failure of this program implies the impossibility of a quantum model. Note that for the purposes of entanglement-assisted communication scenarios, we always choose $p_x=\frac{1}{n_\text{X}}$ and $P_g=\frac{d}{n_\text{X}}$ (where $d$ is the dimension of communication) when communication is classical. When communication is quantum we choose $P_g=\frac{d^2}{n_\text{X}}$.

\section{EA bits are strictly more powerful than hyperbits}	\label{Apphyperbits}

In \cite{Pawlowski2012} an equivalence is claimed between EA bit and hyperbit strategies. While it is true that any hyperbit strategy can be simulated by EA bits, the converse claim is incorrect, as we illustrate with an explicit counterexample. 

We consider the dimension witness of Gallego et al from \ref{sec:gallego}. As discussed there, we found an explicit EA bit strategy achieving $W_5 \approx 9.034$, which we proved to be optimal. The correlations in a hyperbit model are given by the scalar products between unit vectors. Hence, finding the optimal value of a linear function of the correlations for a hyperbit strategy amounts to a search over Gram matrices, which can be cast as a single SDP \cite{blackjack}. Taking into account the possibility that Bob may use more complex strategies, e.g., where he probabilistically decides to discard the message of Alice and outputs a predetermined output), we find, following the methods of \cite{blackjack},  
\begin{equation}
 W_\text{5}\hspace{2mm}\stackrel{\text{hyperbit}}{\leq}\hspace{2mm} 9 \,, 
\end{equation}
which is strictly smaller than the EA bit value. 

The hyperbit construction given in Appendix A of \cite{Pawlowski2012} fails, because the probabilities for Bob to flip his bit, under (eq. A9 \cite{Pawlowski2012}), are not guaranteed to be positive. Taking our optimal EA bit strategy achieving $W_5 \approx 9.034$, we can verify that this is indeed the issue by attempting to transform it into a hyperbit strategy following the construction of \cite{Pawlowski2012}. As expected, the flipping probabilities are negative for most of Bob's inputs and Alice's messages. The explicit calculation can be found in two MATLAB files as ancillary files on the arXiv page of the present paper.

\section{Software Tools}

The code that was used in the numerical analysis of this paper can be found on GitHub: \url{https://github.com/jefpauwels/arXiv-2103.10748-Supplementary-MATLAB-codes}. The repository contains scripts for the information hierarchy as well as the NPA hierarchy for entanglement-assisted classical communication. It also contains  scripts for the heuristic search algorithms we used to find lower bounds for quantum communication, information-restricted communication and entanglement-assisted classical communication, in addition to the optimal strategies achieving the lower bounds quoted in this paper.

\end{document}